\documentclass[prd,aps,groupedaddress,
showpacs,preprintnumbers,
preprint,
nofootinbib,
]{revtex4}%
\usepackage{graphicx}
\usepackage{epsfig}
\usepackage{amssymb}

\def\lQ{\Lambda_{\rm QCD}}
 
\def\als{\alpha_{\rm s}} 
\def\siml{{\ \lower-1.2pt\vbox{\hbox{\rlap{$<$}\lower6pt\vbox{\hbox{$\sim$}}}}\ }}
\def\MS{\overline{\rm MS}}
\newcommand{\be}{\begin{equation}}
\newcommand{\ee}{\end{equation}}
\newcommand{\bea}{\begin{eqnarray}}
\newcommand{\eea}{\end{eqnarray}}
\newcommand{\nn}{\nonumber}

\begin{document}

\title{The QCD static energy at N$^3$LL}
\author{Nora Brambilla}
\affiliation{Physik Department, Technische Universit\"at M\"unchen, D-85748 Garching, Germany}
\affiliation{Dipartimento di Fisica dell'Universit\`a di Milano and INFN, via Celoria 16, 20133 Milan, Italy}
\author{Xavier \surname{Garcia i Tormo}}
\affiliation{High Energy Physics Division, Argonne National Laboratory, 
9700 South Cass Avenue, Argonne, IL 60439, USA}
\author{Joan Soto}
\affiliation{Departament d'Estructura i Constituents de la Mat\`eria and Institut de Ci\`encies del Cosmos, 
Universitat de Barcelona, Diagonal 647, E-08028 Barcelona, Catalonia, Spain}
\author{Antonio Vairo}
\affiliation{Physik Department, Technische Universit\"at M\"unchen, D-85748 Garching, Germany}
\affiliation{Dipartimento di Fisica dell'Universit\`a di Milano and INFN, via Celoria 16, 20133 Milan, Italy}

\date{\today}

\preprint{ANL-HEP-PR-08-57     IFUM-930-FT     UB-ECM-PF 08/03}

\begin{abstract}
We compute the static energy of QCD at short distances at
next-to-next-to-next-to leading-logarithmic accuracy in terms of the three-loop singlet
potential. By comparing our results with lattice data we extract the value of the unknown piece
of the three-loop singlet potential.
\end{abstract}


\maketitle

\section{Introduction} 
The energy between a static quark and a static
antiquark separated at a distance $r$, which is referred to as the
static energy, is a basic object to understand the dynamics of QCD
\cite{Wilson:1974sk}. When  calculated at short distances in
perturbation theory, the virtual emission of ultrasoft gluons
(i.e. gluons with energy and momentum smaller  than $1/r$ that can
change the color state of the quark-antiquark pair  from singlet to
octet) produce infrared divergences.  These, after resummation of
certain diagrams, induce in the  static energy logarithms, $\ln \als
(1/r)$ \cite{Appelquist:1977es}.  In an effective field theory
framework \cite{Pineda:1997bj,Brambilla:1999xf}, the separation of
scales in the problem  is made explicit: the static energy becomes the
sum of a matching coefficient,  the static potential, which encodes
all contributions from the scale $1/r$,   and of the contributions
coming from ultrasoft gluons, which start at three loops. Logarithms
are remnants of the cancellation between infrared divergences of the
potential  and ultraviolet divergences of the ultrasoft contributions
\cite{Brambilla:1999qa}.  These logarithms may be potentially large
when $r$ is very small,  so that $\als(1/r)\ln \als (1/r)$ terms in
the static energy may need to be resummed. The resummation of the
ultrasoft leading  logarithms (LLs)\footnote{Leading ultrasoft
logarithms  contribute to the static energy at order
$\als^{3+n}\ln^{n} \als$ for $n\ge 0$, i.e.  at
next-to-next-to-leading logarithmic, N$^2$LL, order.} was performed in
\cite{Pineda:2000gza}.  The anomalous dimension of the ultrasoft
next-to-leading logarithm (NLL)\footnote{Next-to-leading ultrasoft
logarithms  contribute to the static energy at order
$\als^{4+n}\ln^{n} \als$ for $n\ge 0$, i.e.  at
next-to-next-to-next-to-leading logarithmic, N$^3$LL, order.} was
calculated in \cite{Brambilla:2006wp}\footnote{The calculation relies
on the next-to-leading order (NLO) calculation of the chromoelectric
correlator done in \cite{Eidemuller:1997bb}.} and turned out to be
quite large, showing that the resummation of the ultrasoft NLLs should
also be addressed.  In this paper, we will consider this resummation. 

It has been argued in
\cite{Aglietti:1995tg,Hoang:1998nz,Beneke:1998rk,Pineda:2002se}  that
the proper consideration, and cancellation, of the renormalon
singularities is  crucial to obtain a good convergence of the
perturbative series for the static potential in the short-distance
region. The detailed analysis of the possible influence of ultrasoft
effects in the renormalon structure of the potential will not be
presented in this paper. We will just follow the analysis of
\cite{Aglietti:1995tg,Hoang:1998nz,Beneke:1998rk,Pineda:2002se},
which essentially take advantage of the fact that a constant term may
be added to the potential, a freedom that remains even if ultrasoft
effects are taken into account. As we will see later, this seems to be
enough to obtain a convergent perturbative series for the static
potential in the short-distance region.

The static potential is a basic ingredient of heavy-quarkonium physics
\cite{Brambilla:2004wf}. In particular, its perturbative evaluation at
higher orders is relevant to describe the top-quark pair production
process near threshold. This process is expected to allow the
extraction of the top quark mass to a high precision, and hence a
remarkable effort is being made to calculate it at 
next-to-next-to-next-to-leading order (N$^3$LO)
\cite{Beneke:2008cr,Beneke:2007pj,Beneke:2007gj,Beneke:2005hg}. The
past experience with the next-to-next-to-leading order (N$^2$LO) 
results \cite{Hoang:2000yr} indicates
that both renormalon cancellation and the logarithmic resummation
\cite{Hoang:2002yy,Pineda:2006ri} are necessary for accurate determinations
of the position of the pole and of the shape of the cross section
respectively. Our results will be relevant for the N$^3$LL calculation
of this process.

The paper is structured as follows. In the next section, we introduce
residual mass terms in potential Non-Relativistic QCD (pNRQCD) and
summarize the current status of the perturbative calculations for the
static potential. In section \ref{sec:RenGr}, we present and solve the
renormalization group equations for the static pNRQCD Lagrangian, at
next-to-leading order, and briefly describe the renormalon subtracted
scheme that we use. Section \ref{sec:latt} presents a comparison of
our results for the static energy with lattice data, and the
numerical extraction of the missing piece of the three-loop static
potential. We conclude in section \ref{sec:concl}.

\section{Potentials and residual mass terms in pNRQCD}
The general form of the dimension 6 operators in the pNRQCD Lagrangian is 
\be
c_s \, S^\dagger  S + c_o \, O^{a\, \dagger}O^a\,,
\ee
where $S$ is the singlet and $O^a$ the octet fields.
The coefficients $c_s$ and $c_o$ have dimension 1.

Let us recall that, in order to define Heavy Quark Effective Theory (HQET) beyond perturbation theory, 
or even in perturbation theory when regularizations with an explicit scale (cut-off) are used, 
one needs to introduce a residual mass term $\delta m_Q$ in the Lagrangian \cite{Beneke:1994sw} 
\be
\mathcal{L}_{\rm HQET}=\psi^\dagger \left(iD_0-\delta m_Q\right)\psi 
+\mathcal{O}\left(\frac{1}{m_Q}\right),
\ee
with $m_Q$ the heavy-quark  mass and $\psi$ the heavy-quark field. 
We may associate to $\delta m_Q$ the size of the typical hadronic scale $\lQ$. 
This residual mass term will be inherited by the pNRQCD Lagrangian. 
In the paper, we consider the weak-coupling regime of pNRQCD, defined by 
\be
\label{count}
\frac{1}{r}\gg\frac{\als}{r}\gg\lQ,
\ee
at leading order in the $1/m_Q$ expansion;  
in this situation, the residual mass term is absorbed in the coefficients 
$c_s$ and $c_o$ above. Therefore, it is useful to split them in a part that is
proportional to $1/r$, which  corresponds to the singlet and octet
potential\footnote{The singlet potential is often referred to as the
static potential, a terminology which we also adopt in the paper.
Recall that it coincides with the static energy up to two loops but
differs from it beyond that order.}, and a part that is proportional  to $\lQ$: 
\bea
 c_s &=& V_s+\Lambda_s=
- C_F \frac{\alpha_{V_s}}{r} + \Lambda_s,
\label{eq:cs}  
\\ 
c_o &=& V_o+\Lambda_o=\frac{1}{2N_c} \frac{\alpha_{V_o}}{r} + \Lambda_o, 
\eea
where 
\bea 
&& \hspace{-8mm} 
\alpha_{V_{s,o}}(r,\mu) = \als(1/r)
\Bigg\{1 + \tilde{a}_1\,\frac{\als(1/r)}{4\pi}
\nn
\\ 
&& \hspace{-6mm}  
+ \tilde{a}_{2\,s,o}\,\left(\frac{\als(1/r)}{4\pi}\right)^2  
+ \left[\frac{16\,\pi^2}{3} C_A^3 \,  \ln {r\mu} + \tilde{a}_{3\,s,o}
  \right]\! \left(\frac{\als(1/r)}{4\pi}\right)^3  
\nn
\\ 
&& \hspace{-6mm}  
+ \Bigg[
  a_{4}^{L2}\ln^2 {r \mu}  +  \left(a_{4}^{L}   
  -  \frac{16}{9}\pi^2 \, C_A^3\beta_0 (5 - 6 \ln 2)\right) \ln {r\mu}  
         + \tilde{a}_{4\,s,o}  \Bigg]\! \left(\frac{\als(1/r)}{4\pi}\right)^4 \! +\dots\Bigg\},
\label{VN4LO}
\eea
with
\bea
\hspace{-4mm}
\tilde{a}_1&=&\frac{31}{9}C_A-\frac{10}{9}n_f+2\gamma_E\beta_0, 
\\
\hspace{-4mm}
\tilde{a}_{2,s}&=& \left({4343\over162}+4\pi^2-{\pi^4\over4}+{22\over3}\,\zeta(3)\right)C_A^2
-\left({899\over81}+{28\over 3}\,\zeta(3)\right)C_An_f
\nn
\\
&& -\left({55\over6}-8\,\zeta(3)\right)C_Fn_f +\left({10\over9}n_f\right)^2
\nn
\\
&& +\left(\frac{\pi^2}{3}-4\gamma_E^2\right)\beta_0^2
  +\gamma_E\left(4\tilde{a}_1\beta_0+2\beta_1\right),
\\
\hspace{-4mm}
\tilde{a}_{2,o} &=& \tilde{a}_{2,s} + C_A^2(\pi^4 -12\pi^2),
\\
\hspace{-4mm}
a_{4}^{L2}&=& \frac{16\pi^2}{3}C_A^3\left(-\frac{11}{3}C_A+\frac{2}{3}n_f\right),
\\
\hspace{-4mm}
a_4^L &=&  16\pi^2C_A^3\left[\tilde{a}_1
+ n_f \left( -\frac{20}{27} + \frac{4}{9} \ln 2\right)
+ C_A\left(\frac{149}{27}-\frac{22}{9}\ln 2+\frac{4}{9}\pi^2\right)\right];
\eea
$\mu$ is the ultrasoft factorization scale. The color factors are defined as 
$C_F= (N_c^2-1)/(2N_c)$, $C_A=N_c$ where $N_c$ is the number of colors;
$n_f$ is the number of (massless) flavors; $\gamma_E$ is the Euler constant.
The strong coupling constant $\als$ is in the $\MS$ scheme. The beta function is defined as
\be
\als\beta(\als) = \frac{d\,\als}{d\ln \mu} = -\frac{\als^2}{2\pi}
\sum_{n=0}^\infty \left( \frac{\als}{4\pi} \right)^n \beta_n, 
\ee
where $\beta_0 = 11 C_A/3 - 2 n_f/3$, 
$\beta_1 = 34 C_A^2/3 - 10 C_A  n_f/3 - 2 C_F  n_f$, and explicit expressions of 
$\beta_2$ and $\beta_3$ may be found, for instance, in \cite{vanRitbergen:1997va,Czakon:2004bu}.

The one-loop coefficient $\tilde{a}_1$ was calculated in \cite{Fischler:1977yf,Billoire:1979ih}, 
the two-loop singlet coefficient $\tilde{a}_{2,s}$ 
in \cite{Peter:1996ig,Peter:1997me,Schroder:1998vy,Kniehl:2001ju} 
and the two-loop octet coefficient $\tilde{a}_{2,o}$ in \cite{Kniehl:2004rk}.
The logarithmic piece of the third-order correction was calculated in
\cite{Brambilla:1999qa,Kniehl:1999ud,Brambilla:1999xf,Hoang:2002yy}, 
whereas the non-logarithmic piece $\tilde{a}_{3\,s,o}$ 
has not been completely calculated yet. The fermionic contributions of
$\tilde{a}_{3\,s}$ has been presented very recently in
\cite{Smirnov:2008pn}, where the computation of the $n_f$ independent
piece is reported to be in progress. A Pad\'e estimate of $\tilde{a}_{3,s}$ gives:
$\tilde{a}_{3,s}=-48\pi^3\, V_{s}^{(3)}\,$, $V_{s}^{(3)}(n_f=3)=-38.4$, 
$V_{s}^{(3)}(n_f=4)=-28.7$, $V_{s}^{(3)}(n_f=5)=-20.5$  \cite{Chishtie:2001mf}. The
double logarithmic coefficient $a_4^{L2}$ may be obtained from
\cite{Pineda:2000gza,Brambilla:2006wp} and the logarithmic coefficient $a_4^L$ was
obtained in \cite{Brambilla:2006wp}\footnote{Only the coefficient of the
singlet potential was obtained there, it will be shown later in the paper
that it coincides with the coefficient of the octet potential.}. 
$\Lambda_{s,o}$ stands for $\Lambda_{s,o}(r,\mu)$\footnote{
To simplify the notation, we will often suppress 
the dependence on $r$ and just write $\Lambda_{s,o}(\mu )$.}.

At order $r^0$ in the multipole expansion, the dimension 6 operators
of pNRQCD do not have an anomalous dimension and, therefore, the
renormalization group equations for the coefficients $\Lambda_s$ and
$\Lambda_o$ will have the same structure as in the HQET the renormalization 
group equation for the coefficient of the operator $\psi^\dagger\psi$ has. 
At next-to-leading order in the multipole expansion, the pNRQCD Lagrangian 
reads 
\bea
{\mathcal L}_{\rm pNRQCD} &=& 
{\mathcal L}_{\rm light} 
+ \int d^3{\bf r} \; {\rm Tr} \,  \Biggl\{ {\rm S}^\dagger \left[ i\partial_0 - c_s(r,\mu ) \right] {\rm S} 
+ {\rm O}^\dagger \left[ iD_0 - c_o(r,\mu ) \right] {\rm O} \Biggr\}
\nn\\
&& 
+ V_A ( r, \mu) {\rm Tr} \left\{  {\rm O}^\dagger {\bf r} \cdot g{\bf E} \,{\rm S}
+ {\rm S}^\dagger {\bf r} \cdot g{\bf E} \,{\rm O} \right\} 
\nn\\
&&
+ {V_B (r, \mu) \over 2} {\rm Tr} \left\{  {\rm O}^\dagger {\bf r} \cdot g{\bf E} \, {\rm O} 
+ {\rm O}^\dagger {\rm O} {\bf r} \cdot g{\bf E}  \right\}
+ \dots ,
\label{pNRQCD}
\eea
where ${\mathcal L}_{\rm light}$ is the part of the Lagrangian involving
gluons and light quarks, which coincides with the QCD one, 
${\rm S} =  1\!\!{\rm l}_c/\sqrt{N_c} S$, 
${\rm O} = \sqrt{2} T^a O^a$, ${\bf E}$ is the chromoelectric field, 
$V_A$ and $V_B$ are matching coefficients associated with the 
$\mathcal{O}(r)$ operators of the pNRQCD Lagrangian and 
the dots stand for higher-order terms in the multipole expansion.
Ultrasoft gluons cause transitions between singlet and octet fields and generate an
ultrasoft anomalous dimension for the dimension 6 operators. In particular, 
this modifies the renormalization group (RG) equations for $\Lambda_s$ and $\Lambda_o$.

\section{Renormalization Group}
\label{sec:RenGr}
The general structure of the renormalization group equations of pNRQCD
in the static case has been discussed in \cite{Pineda:2000gza}, where
the complete N$^2$LL order was calculated. Here, we will calculate the 
complete N$^3$LL order. It has been proved in \cite{Brambilla:2006wp} that in order to
perform the calculation one needs not to consider higher orders in the
multipole expansion beyond those already contributing to the N$^2$LL
calculation. Hence, the structure of the RG equations remains the same
as in \cite{Pineda:2000gza}, but the anomalous dimensions need to be
calculated to one order more in the ultrasoft loops. The RG equations read
\bea
\left\{
\begin{array}{l}
\mu \displaystyle {d\over d  \mu }  c_{s}  = \gamma_s (\als )V_A^2 
 \left( c_o-c_s  \right)^3 r^2
\\ \\
\mu \displaystyle {d\over d  \mu }  c_{o}  = \gamma_o (\als )V_A^2 
 \left( c_o-c_s  \right)^3 r^2
\\ \\
 \mu  \displaystyle {d\over d  \mu}  \als    =  \als  \beta(\als)\\
\\
 \mu  \displaystyle {d\over d  \mu }  V_A   = \gamma_A (\als ) V_A\\
\\
 \mu  \displaystyle {d\over d  \mu }  V_B   = \gamma_B (\als ) V_B\\
\end{array}
\right. \,,
\label{rg}
\eea 
where the anomalous dimensions $\gamma_s (\als )$,  $\gamma_o (\als )$, $\gamma_A (\als )$ 
and $\gamma_B (\als )$ are needed at order $\als^2$. 
Strictly speaking the equations above hold for $c_{s,o}=V_{s,o} + \Lambda_{s,o}$, 
provided that $V_{s,o} \gg \Lambda_{s,o}$ and one stays at linear order in $\Lambda_{s,o}$. 
If quadratic or cubic terms in $\Lambda_{s,o}$ are included, 
additional counterterms in the potential are needed to absorb the ultraviolet divergences 
of the ultrasoft calculation. 

\subsection{Anomalous dimensions}
We calculate the anomalous dimensions in a regularization scheme in
which the gluons, light quarks and center of mass motion are taken in $D$ dimensions 
but the potentials in the ultrasoft loops 
are kept in three dimensions. We renormalize using the 
$\overline{\rm MS}$ scheme in (relative) coordinate space.

In this scheme, the anomalous dimension $\gamma_s$ was obtained in
\cite{Brambilla:2006wp}. It is $-2\, \als \,$ $ \times \partial Z^{(1)}/\partial \als$, 
where $Z^{(1)}$ denotes the coefficient of the $1/\hat{\epsilon}$ 
($1/\hat{\epsilon}=1/\epsilon-\gamma_E+\ln 4\pi$, $D=4-2\epsilon$) 
pole of the ultrasoft contribution to the static energy; we have 
\be 
\gamma_s(\als )=-\frac{2}{3}\frac{ \als C_F}{\pi}  
\left( 1 +6\, \frac{ \als}{\pi}\,B\right), 
\ee 
with
\be
B = \frac{-5n_f + C_A(6\pi^2+47)}{108}.
\ee

\begin{figure}
\centering
\includegraphics[width=8cm]{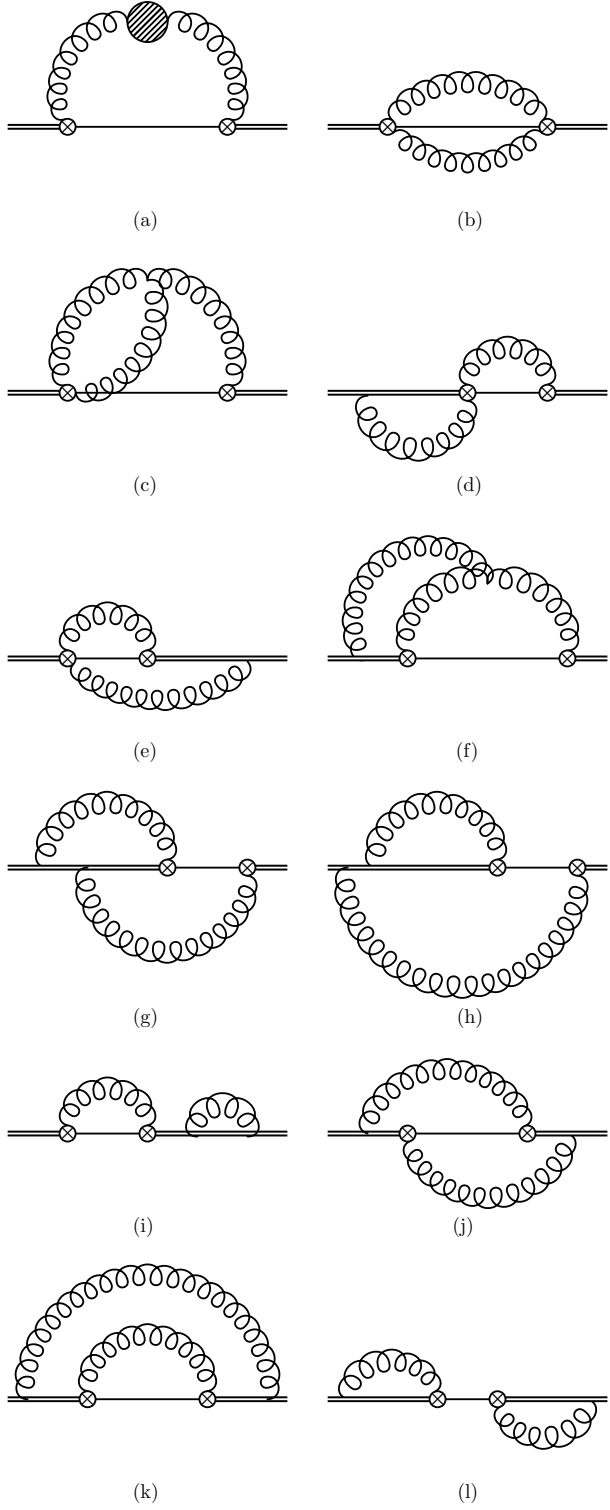}
\caption{Ultrasoft contributions to the self-energy of the octet field  in a covariant gauge, 
at the order of interest. Symmetric graphs are understood for (c)-(i).}\label{VONLO}
\end{figure}

In a similar way, $\gamma_o$ can be obtained from the $1/ \hat{\epsilon}$  
poles of the ultrasoft contribution to the self-energy of
the octet field. 
It is important to recall that, although we may obtain it by matching 
gauge-dependent Green's functions, the self-energy of the
octet field is a gauge invariant quantity in perturbation theory, for the same
reason as the pole mass is. At the order we are interested in, it can
be obtained from the following expression to be taken in the $T\to\infty$ 
limit \cite{Brambilla:1999xf}
\bea
\langle T^a W_\Box T^b \rangle 
&=& Z_o(r) 
e^{-iTV_o(r)} \Bigg( \langle \phi(T/2,-T/2)_{ab}^{\rm adj} \rangle   
\label{vopnrqcdus}
\nn\\ 
& & 
- { g^2 \over 2N_c} V_A^2 (r) \int_{-T/2}^{T/2} \!\! dt 
\int_{-T/2}^{t} \! dt^\prime e^{-i(t-t^\prime)(V_s-V_o)} 
\nn\\
&& 
\times\langle \phi(T/2,t)^{\rm adj}_{aa^\prime} {\bf r}\cdot {\bf E}^{a^\prime}(t) 
{\bf r}\cdot {\bf E}^{b^\prime}(t^\prime) \phi(t^\prime,-T/2)^{\rm adj}_{b^\prime b}\rangle\Bigg). 
\eea
$T^a W_\Box T^b$ stands for a $T^a$ and a $T^b$ insertion 
at the time $T/2$ and $-T/2$ respectively in the space sides of a rectangular Wilson loop, and 
$\phi (t,t')^{\rm adj}_{ab}$ is a Wilson line in the adjoint
representation.  The Feynman diagrams involved in the evaluation of
this quantity in a covariant (or Coulomb) gauge do not coincide with
the ones needed for $\gamma_s$ (compare Fig. \ref{VONLO} with Fig. 4 of
\cite{Brambilla:2006wp}), and would require extra
calculations. Fortunately there is an argument which makes the
explicit calculation unnecessary. If we take the $A_0=0$ gauge, the
number of diagrams to be evaluated collapses to a few (the octet field
does not emit gluons anymore), which, in addition, are the same for
$\gamma_s$ and $\gamma_o$. In this gauge, both anomalous dimensions are
related by trivial color factors. Since both anomalous dimensions are
gauge invariant, it turns out that we can read $\gamma_o$ from the
known result for $\gamma_s$:
\be 
\gamma_o(\als ) =-{\gamma_s(\als)\over N_c^2-1}.  
\ee 
At order $\als$, this is confirmed by the explicit
calculation  of \cite{Brambilla:1999xf}. 

\begin{figure}
\centering
\includegraphics{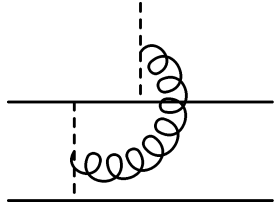}
\caption{Infrared divergent diagram in the matching calculation of $V_A$ and $V_B$.}\label{VANLO}
\end{figure}

Let us turn now to the evaluation of $\gamma_A$ and $\gamma_B$. 
We use the fact that the $\mu$ dependence of $V_A$ and $V_B$ can be also 
obtained from the infrared logarithms in the matching calculation
between HQET and pNRQCD. There is only one diagram which is infrared 
divergent in the matching calculation: it is displayed in
Fig. \ref{VANLO}. However, the divergence turns out to be linear and
produces no logarithms. Then these anomalous dimensions remain zero also at
next-to-leading order:      
\be 
\gamma_A(\als ) =\gamma_B(\als ) =0.  
\ee

\subsection{Solution of the renormalization group equations}
The renormalization group equations for $V_s$ and
$V_o$ can be read from (\ref{rg}) using that $V_{s,o}
\gg \Lambda_{s,o}$ and neglecting the latter. They are given by
\bea
\left\{
\begin{array}{l}
\mu \displaystyle {d\over d  \mu }  V_{s}  = 
-{2 \over 3}\frac{ \als C_F}{\pi} \left( 1 +6\, \frac{ \als}{\pi}\,B \right)V_A^2 
 \left( V_o-V_s\right)^3 r^2
\\
\\
 \mu  \displaystyle {d\over d  \mu }  V_{o}  = 
{1\over 3}\frac{ \als}{\pi} \frac{1}{N_c}\left( 1 +6\, \frac{ \als}{\pi}\,B \right)V_A^2 
 \left( V_o-V_s\right)^3 r^2
\\
\\
 \mu  \displaystyle {d\over d  \mu}  \als    =  \als  \beta(\als)
\\
\\
 \mu  \displaystyle {d\over d  \mu }  V_A   = 0 
\\
\\
 \mu  \displaystyle {d\over d  \mu }  V_B   = 0 
\\
\end{array}
\right. \,.
\eea
The solutions of these equations, at the order we are interested in, are $V_A (\mu)=V_A (1/r)=1$, 
$V_B(\mu)=V_B(1/r)=1$ (from $V_{A,B}(1/r) = 1+{\cal O}(\als^2)$, see \cite{Brambilla:2006wp}), and 
\bea
V_{s}(\mu)  &=&
V_{s}(1/r)  
+\frac{2}{3} C_F r^2 \left[V_o(1/r)-V_s(1/r)\right]^3
\nn\\
&&
\quad\quad\quad\quad\quad\quad\quad
\times
\left( \frac{2}{\beta_0}  \ln \frac{\als(\mu)}{\als(1/r)} 
+ \eta_0\left[\als(\mu)- \als(1/r) \right] \right) \,,\label{eq:VsN3LL}
\\
V_{o}(\mu)  &=&
V_{o}(1/r)  
-\frac{1}{3 N_c}  r^2 \left[V_o(1/r)-V_s(1/r)\right]^3
\nn\\
&&
\quad\quad\quad\quad\quad\quad\quad\times
\left( \frac{2}{\beta_0}  \ln \frac{ \als(\mu)}{\als(1/r)} 
+ \eta_0 \left[ \als(\mu) - \als(1/r) \right] \right) \,, 
\label{alphaVso}
\eea
where 
\be
\eta_0 = \frac{1}{\pi}\left(-\frac{\beta_1}{2\beta_0^2} + \frac{12B}{\beta_0}\right).
\ee

The renormalization group equations for $\Lambda_s$ and 
$\Lambda_o$ can also be obtained from (\ref{rg}) by expanding $c_{s,o}$ about $V_{s,o}$ 
and keeping the terms linear in $\Lambda_{s,o}$. They are given by
\bea
\hspace{-0.7cm}
\left\{
\begin{array}{l}\label{eq:difeql}
\mu \displaystyle {d\over d  \mu }  \Lambda_s  = 
-2\frac{ \als C_F}{\pi} \left( 1 +6\, \frac{ \als}{\pi}\,B \right)V_A^2r^2
\left[V_o(1/r)-V_s(1/r)\right]^2 (\Lambda_o-\Lambda_s)
\\
\\
 \mu  \displaystyle {d\over d  \mu }  \Lambda_o  = \frac{ \als}{\pi} \frac{1}{N_c}
\left( 1 +6\, \frac{ \als}{\pi}\,B \right)
V_A^2r^2\left[V_o(1/r)-V_s(1/r)\right]^2 (\Lambda_o-\Lambda_s)
\\
\\
\mu  \displaystyle {d\over d  \mu}  \als    =  \als  \beta(\als)
\end{array}
\right. \!\!,
\eea
where we have already approximated $V_o(\mu)$ and $V_s(\mu)$ by $V_o(1/r)$ and $V_s(1/r)$
(the $\mu$ dependence of  $V_o(\mu)$ and $V_s(\mu)$ enters at N$^3$LO, 
which is beyond the accuracy of (\ref{eq:difeql})).
The solutions of the renormalization group equations read, 
\bea
\Lambda_s(\mu)  &=& N_s \Lambda + 2\,C_F(N_o-N_s)\Lambda\, r^2\left[V_o(1/r)-V_s(1/r)\right]^2
\nn\\
&&
\quad\quad\quad\quad\quad\quad\quad
\times\left(\frac{2}{\beta_0}\ln {\als(\mu)\over \als (1/r)} + \eta_0 \left[\als (\mu)-\als (1/r)\right]\right)\!, 
\label{lambdas}\\
\Lambda_o(\mu)  &=& N_o \Lambda - \frac{1}{N_c}(N_o-N_s)\Lambda\,
 r^2\left[V_o(1/r)-V_s(1/r)\right]^2
\nn\\
&&
\quad\quad\quad\quad\quad\quad\quad\times
\left(\frac{2}{\beta_0}\ln {\als(\mu)\over \als (1/r)} + \eta_0 \left[\als(\mu)-\als (1/r)\right]\right), 
\label{lambdao}
\eea
where $N_s$, $N_o$ are two arbitrary scale-invariant dimensionless constants and  
$\Lambda$ is an arbitrary scale-invariant quantity of dimension one. 

The integration constants $N_s$ and $N_o$ are fixed by the initial 
conditions, $\Lambda_s(1/r)$ and $\Lambda_o(1/r)$, of the solutions
of the RG equations. In turn, the initial conditions are fixed by matching 
a suitable Green's function in QCD with the corresponding one in pNRQCD. 
Note that if at the matching scale $\Lambda_s (1/r)=\Lambda_o (1/r)=2\delta m_Q$, 
as it happens in  MS-type schemes, then $\Lambda_s (\mu)=\Lambda_o (\mu)=2\delta m_Q$ for any $\mu$. 
We will see, in the following sections, the convenience to use an RS (renormalon subtracted) scheme. 
In the RS scheme $\Lambda_{s}(1/r )$ and $\Lambda_{o}(1/r )$ are different constants 
(which also differ from $2\delta m_Q$) and hence $\Lambda_s (\mu)$ and $\Lambda_o (\mu)$ 
evolve in a non-trivial way according to the RG equations above.

\subsection{The Renormalon Subtracted scheme}
\label{subsec:RS} 
The discussion in the previous sections is independent of the
renormalization scheme used for the matching calculation between HQET
and static pNRQCD at the scale $1/r$. The outcome of the matching
calculation only enters through the initial conditions of the RG
equations. It is well known that the singlet potential $V_s$
calculated in the $\overline{\rm MS}$ scheme displays a bad behavior
as a series in $\als (1/r)$ even at small values of $r$. This bad
behavior may be ascribed to renormalon singularities that lie very
close to the origin of the Borel plane. In order to treat the
renormalon singularity, we shall follow the procedure described in
Ref. \cite{Pineda:2001zq}, the so-called renormalon subtracted (RS)
scheme. Under RS scheme we understand a class of subtraction schemes
that subtract from the  perturbative series of $V_{s,o}$ in the $\MS$
scheme the non-integrable piece  at $u=1/2$ in the Borel transform of
the potential, expanded about $u=0$ and integrated over $u$.  The
whole non-analytic piece (non-integrable and integrable) at $u=1/2$,
expanded about $u=0$ and integrated over $u$, reads (at the scale $\rho$) 
\bea  
R_{s,o} \, \rho \,  \sum_{n=1}^\infty \left(
\frac{\beta_0}{2\pi} \right)^n \als(\rho)^{n+1} \sum_{k=0}^\infty
d_{k} \frac{\Gamma(n+1+b-k)}{\Gamma(1+b-k)}\,.
\label{RSfullscheme}
\eea

The coefficients $d_k$ are given in terms of the coefficients of the beta function. 
Since the beta function is known up to four loops only, all $d_k$ for $k\ge3$ are unknown; 
the known terms are
\bea
d_0 & = & 1\, ,\nn\\
d_1 & = & \frac{\beta_1^2-\beta_2\beta_0}{4 b
   \beta_0^4}\, ,\nn\\
d_2 & = & \frac{-2 \beta_0^4 \beta_3+4 \beta_0^3
   \beta_1 \beta_2+\beta_0^2
   \left(\beta_2^2-2 \beta_1^3\right)-2 \beta_0 \beta_1^2 \beta_2+\beta_1^4}{32 (b-1) b
   \beta_0^8}\, ,
\eea
with
\begin{equation}
b=\frac{\beta_1}{2\beta_0^2}\, .
\end{equation} 
Hence, it is practically unfeasible to subtract
Eq. (\ref{RSfullscheme}). This is not a real problem  because only
subtracting the $k=0$ term, which corresponds to the non-integrable
piece in the Borel transform, is necessary in order to obtain a series
that is  Borel summable. In the following, we will subtract all the
known terms in Eq. (\ref{RSfullscheme}), i.e. up to $k=2$, as was done
in the original proposal of the RS \cite{Pineda:2001zq}.

Furthermore, since the potential is given as an expansion of $\als
(1/r)$,  in order to achieve a successful renormalon subtraction at
every order in $\als$,  it is important to expand $\als(\rho)$ in
terms of $\als(1/r)$ (or viceversa).  We chose to expand $\als(1/r)$
in terms of $\als(\rho)$ in Eq. (\ref{VN4LO})  instead of doing the
reverse in Eq. (\ref{RSfullscheme}), because the uncertainty in the
normalization constants $R_{s,o}$ of the renormalon singularities is
then largely absorbed in the arbitrary additive constant needed to
compare with lattice data (as it will be described in the next
section). This expansion generates $\ln r\rho$ terms, which will be
kept from becoming large by choosing
$\rho=1/\langle r \rangle =3.25/r_0$, $\langle r \rangle$ being the central
value of the range where we compare with lattice data and $r_0$ being the
reference scale used in the lattice computation (see the next section). At
the order we are working, we only need to keep terms up to order
$\als^4 (\rho)$.

\section{Comparison with lattice results}
\label{sec:latt}
In this section we will compare our results with the ($n_f=0$) lattice data 
of Ref. \cite{Necco:2001xg}. This will allow us to extract a value for 
the three-loop coefficient $\tilde{a}_{3\,s}$.

\subsection{Setting the scales and parameters}
\label{subsec:scalesandpar}
We choose, as anticipated in the previous section, $\rho=1/\langle r \rangle = 3.25/r_0$ 
(the reference scale $r_0$, used in the lattice computation, has a value of about 0.5 fm, 
see \cite{Necco:2001xg} for more details; we will present all our results
in units of $r_0$). The remaining scales and parameters entering in
the expressions are chosen as follows. The number of light flavors
$n_f$ is set to zero. The ultrasoft scale $\mu$ is set to
$\mu=2/r_0$. The normalizations of the $u=1/2$ renormalon
singularities for the singlet and octet potentials, $R_{s,o}$, 
are determined using the procedure described in \cite{Lee:1999ws};  
one obtains:
\be
\label{eq:RsRo}
\begin{array}{ccccc}
R_s & = & -1.333+0.499-0.338 & = & -1.172,
\\
R_o & = & 0.167-0.0624+0.00972 & = & 0.114.
\end{array}
\ee
$\als$ at the relevant scales is determined according to \cite{Capitani:1998mq}, 
which uses $\Lambda_{\MS}\,r_0=0.602(48)$. We will use the running of $\als$ 
according to the order we are working at (for instance, for the one-loop curve we use 
the two-loop running for the order $\als$ term and the one-loop running for the order $\als^2$ term, and so on).

The three-loop coefficient $\tilde{a}_{3,s}$, which enters in our N$^3$LL results, is unknown. 
We can use the Pad\'e estimate of \cite{Chishtie:2001mf} to get an idea of its expected size
(that is, use $c^{\rm Pred}_{0,n_f=0}=313$ from Table 1 of that paper, which corresponds to 
$\tilde{a}_{3,s} = 114633$; the relation between $c_0$, $V_s^{(3)}$ 
and ${\tilde a}_{3,s}$ is given in Eq. (29d) of \cite{Chishtie:2001mf}, 
for easier reference we reproduce it in the appendix). 
To
estimate the uncertainty that we should associate to that Pad\'e value we can make use of the results of
\cite{Smirnov:2008pn}. In Ref. \cite{Smirnov:2008pn}, all the fermionic
contributions to the three-loop coefficient $c_0$ are calculated, therefore the difference
\be
\label{eq:diffas}
c_{0,n_f}-c_{0,n_f-1}
\ee
is known ($c_{0,n_f}$ stands for the coefficient $c_0$ calculated with $n_f$
flavors). Since Ref.~\cite{Chishtie:2001mf} presents the Pad\'e
estimated values of $c_0$ for $n_f=0,\dots,6$, we can check if
those results satisfy the known values for (\ref{eq:diffas}) or not. This
comparison is presented in Table \ref{tab:diffs}\footnote{In \cite{Smirnov:2008pn}, 
the coefficient of the $d_F^{abcd}d_F^{abcd}$ color structure is given numerically, 
but the limited numerical precision is not yet affecting the numbers presented in Table~\ref{tab:diffs}.}. 
In addition, we can also obtain $c_0$ for $n_f=0$, which is the coefficient we
need here, from the Pad\'e estimated value of $c_0$
for $n_f=6$ and (\ref{eq:diffas}):\footnote{We chose $n_f=6$, because it is for $n_f=6$ 
that the Pad\'e approximation comes closest to the three-loop
RG accessible coefficient $c_1$, in the notation of
\cite{Chishtie:2001mf} that we reproduce in Eqs. (\ref{eq:Vs3}) and (\ref{eq:a3Vs3}).}
\be
\label{eq:c0from6}
c_{0,n_f=0}^{{\rm from}\,n_f=6}=239,
\ee
to be compared with the value $c_{0,n_f=0}=313$ presented in \cite{Chishtie:2001mf}. 
We will take this as an indication that one should assign an uncertainty of around $30-40\%$ 
to the Pad\'e estimate. As we will see later, though, the lattice data is precise enough to be used 
to obtain an independent extraction of the value of $c_0$. 

\begin{table}
\begin{center}
\begin{tabular}{c|c|c}
 & Exact result \cite{Smirnov:2008pn} & Pad\' e estimate \cite{Chishtie:2001mf}\\
\hline
$n_f=6$ & -21.39 & -29.6 \\
\hline
$n_f=5$ & -26.56 & -37.4 \\
\hline
$n_f=4$ & -31.86 & -44.5 \\
\hline
$n_f=3$ & -37.28 & -51 \\
\hline
$n_f=2$ & -42.84 & -57 \\
\hline
$n_f=1$ & -48.52 & -63 \\
\end{tabular}
\caption{Values of the difference $c_{0,n_f}-c_{0,n_f-1}$ for the
  exact result and the Pad\'e estimates.}\label{tab:diffs}
\end{center}
\end{table}

\subsection{The static potential} 
\begin{figure}
\centering
\includegraphics[width=12cm]{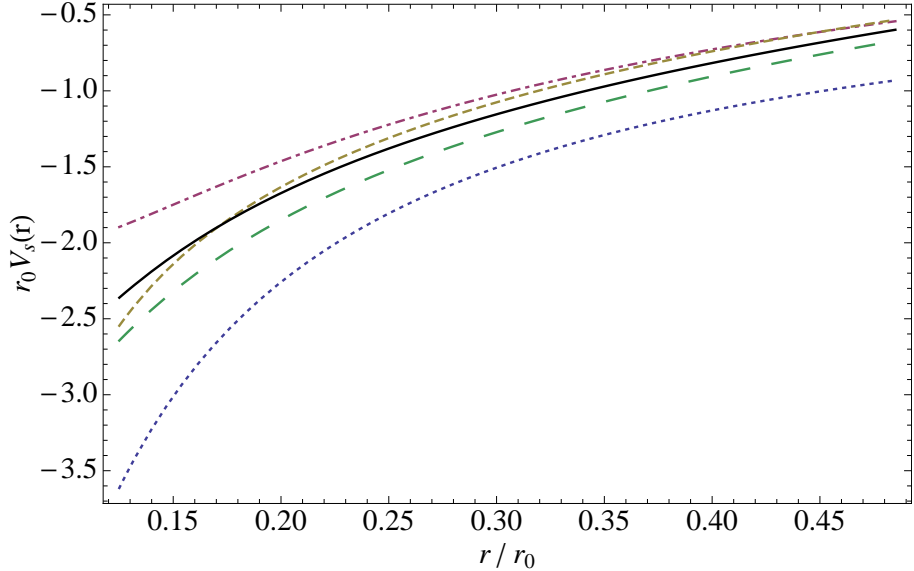}
\caption{Static potential $r_0V_s(r)$, in the RS scheme, as a function
  of $r/r_o$. The dotted blue curve is at tree level, the dot-dashed
  magenta curve is at one loop, the dashed brown curve is at two loop plus 
  leading ultrasoft logarithmic resummation and the long-dashed green curve is
  at three loop (Pad\' e estimate) plus next-to-leading ultrasoft
  logarithmic resummation. The solid black curve is also at three-loop plus next-to-leading ultrasoft
  logarithmic resummation but using $c_0=250.7$, see section \ref{subsec:exc0}.}
\label{fig:Vsnous}
\end{figure}

Using the choices of scales and
parameters described in the previous section, we obtain the singlet
potential in the RS scheme represented in Fig. \ref{fig:Vsnous} (to
alleviate the notation we do not explicitly indicate the dependences
on $\mu$ and $\rho$ of the different functions in the labels of the
plots). In all the plots in this section, the dotted
blue curve will be at tree level, the dot-dashed magenta curve will be
at one loop, the dashed brown curve will be at two loop plus N$^2$LL
resummation and the long-dashed green curve will be at three loop
(with the Pad\'e estimated value for $c_0$, i.e. $c_0=313$) plus
N$^3$LL resummation. As expected and in sharp contrast to what would happen 
in an on-shell scheme \cite{Pineda:2002se}, we see that when we use a threshold scheme 
that cancels the leading renormalon, like the RS scheme, 
the perturbative series for the potential exhibits a convergent behavior. 

\subsection{The static energy}
In order to perform a comparison with lattice data, 
we have to plot the static energy $E_0$ as a function of $r$:
\be
\hspace{-7mm}
E_0(r)=V_s+\Lambda_s+\delta_{\rm US}=V_s\left(r,\mu,\rho\right)
+K_1(\rho)+K_2(\rho)f(r,\mu,\rho)+\delta_{\rm US}\left(r,\mu\right),
\label{eq:Vpf}
\ee
where $\Lambda_s$ is given by Eq. (\ref{lambdas}) and $\delta_{\rm US}$ 
contains the contributions from ultrasoft gluons. $V_s$, $K_1$ and $K_2$ 
have to be understood in the RS scheme, which 
is where the $\rho$ dependence comes from (at the order we are working, 
$f$ will not depend on $\rho$, which will be dropped from it in the following). 
Also, at the order we are working, the renormalized expression of $\delta_{\rm US}$ 
is only needed at leading order\footnote{
Note that large logarithms have been resummed in $V_s$, so that
the counting of $\delta_{\rm US}$ is a fixed order one if $\mu\sim V_o-V_s $.} 
(and can be read from equation (14) of \cite{Brambilla:2006wp}):
\be
\label{eq:deltaus}
\delta_{\rm US}=C_F\frac{C_A^3}{24}\frac{1}{r}\frac{\als(\mu)}
{\pi}\als^3(1/r)\left(-2\ln \frac{\als(1/r)N_c}{2r\,\mu}+\frac{5}{3}-2\ln 2 \right).
\ee
$K_1$ and $K_2$ are the constants 
$N_s\Lambda$ and $(N_o-N_s)\Lambda$ in Eq. (\ref{lambdas}) while $f(r,\mu)$ is
\bea
f(r,\mu)=2C_F r^2\left[V_o(1/r)-V_s(1/r)\right]^2
\left(\frac{2}{\beta_0}\ln {\als(\mu)\over \als (1/r)} + \eta_0 \left[\als(\mu)-\als (1/r)\right]\right).
\nn\\
\label{eq:f}
\eea

\begin{figure}
\centering
\includegraphics[width=12cm]{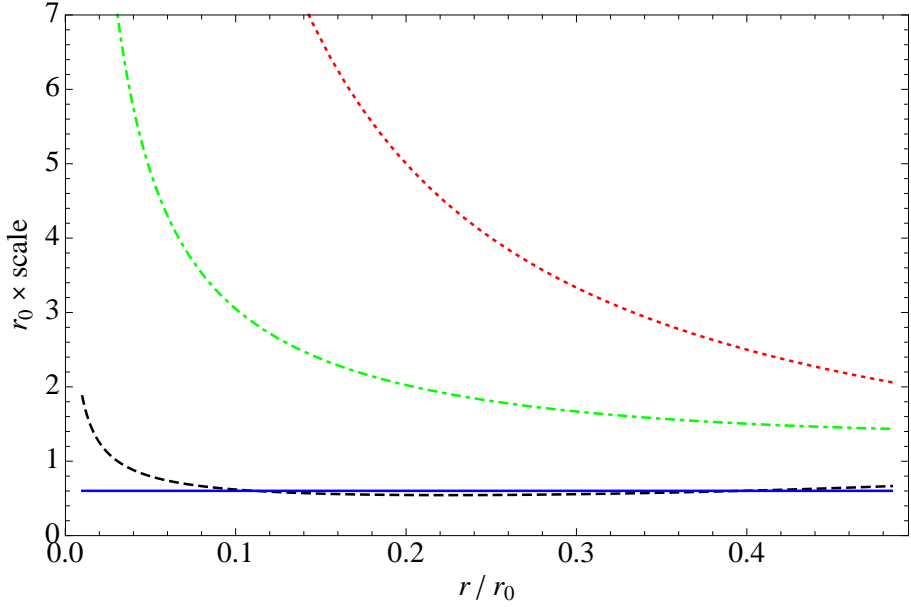}
\caption{Illustration of the hierarchy of scales relevant for the static energy. 
The scales are: $1/r$ (dotted red), $(V_o-V_s)$ (dot-dashed green), $\als(V_o-V_s)$ (dashed black) 
and $\Lambda_{\MS}=0.602/r_0$ (solid blue). $(V_o-V_s)$ is taken at tree level, 
$(V_o-V_s)={N_c\als}/{(2r)}$, and the one-loop running of $\als=\als(1/r)$ is used. \label{fig:pc}}
\end{figure}

We are considering the weak-coupling regime in the static limit, 
defined by the hierarchy of scales given in (\ref{count}). 
To have a definite way to organize the terms in Eq. (\ref{eq:Vpf}) we will also use 
\be
\Lambda\sim N_s\Lambda\sim N_o\Lambda\sim\lQ\sim\frac{\als^2}{r},
\ee
which is compatible with Eq. (\ref{count}). Figure \ref{fig:pc} shows that the scale 
hierarchy of Eq. (\ref{count}) as well as the counting above hold for $r/r_0 \le 0.5$.

The lattice data of \cite{Necco:2001xg} is presented as the difference between the static 
energy at distance $r$ and the static energy at a reference scale $r_c$. 
Therefore, what we actually need to plot, in order to compare with the lattice data, is
\begin{equation}
\label{eq:Etilde}
E_0(r)-E_0(r_{\rm min})+E_0^{\rm latt.}(r_{\rm min})=V_s+\tilde{K}_1+K_2f+\delta_{\rm US},
\end{equation}
where $r_{\rm min}$ is the shortest distance at which lattice data is available, 
 $r_0E_0^{\rm latt.}(r_{\rm min})=-1.676$ is given in Table 2 of \cite{Necco:2001xg}, and
 ${\tilde K}_1$ is a suitable constant, which can be obtained by imposing
 the equation to be true at $r=r_{\rm min}$.

From the counting described above, we see that the last two terms in the rightmost part 
of Eq. (\ref{eq:Etilde}) only start contributing at the three-loop level. 
We also see that, at three-loop level, we only need the function $f(r)$ 
in Eq. (\ref{eq:f}) at leading order, i.e.
\be
\label{eq:fLO} 
f(r,\mu)=2C_F \left(\frac{\als(1/r)}{2N_c}+C_F\als(1/r)\right)^2\frac{2}{\beta_0}\ln {\als(\mu)\over \als (1/r)}.  
\ee 
Therefore, the static energy (that we will use to compare with lattice data) 
at N$^3$LL order is given by Eq. (\ref{eq:Etilde}) with $V_s$ given by (\ref{eq:VsN3LL})
(understood in the RS scheme), $f$ given by (\ref{eq:fLO}) and
$\delta_{\rm US}$ given by (\ref{eq:deltaus}). The constant $K_2$ will
be fixed by a fit to the lattice data below $r/r_0=0.5$. In all the
plots, we will always display our results until $r/r_0=0.5$,  which is
the region where we expect perturbation theory and our hierarchy of
scales to be reliable.

\subsubsection{Analysis of the uncertainties in the static energy}
\label{subsubsec:uncpot} 
As we have already mentioned, our N$^3$LL results depend on the unknown 
three-loop coefficient $c_0$ (in addition to the constant $K_2$). 
The static energy also suffers from uncertainties due to the $\Lambda_{\MS}$ parameter, used
to determine $\als$, and from uncertainties due to the neglecting of $\als^5/r$ and higher-order 
terms. Since we want to use the lattice data to extract $c_0$, 
we need to make sure that the static energy is more sensitive to variations 
of $c_0$ (and $K_2$, which will be also fitted to the data) than to the errors 
due to $\Lambda_{\MS}$ and the higher-order terms.

\begin{figure}
\centering
\includegraphics[width=8cm]{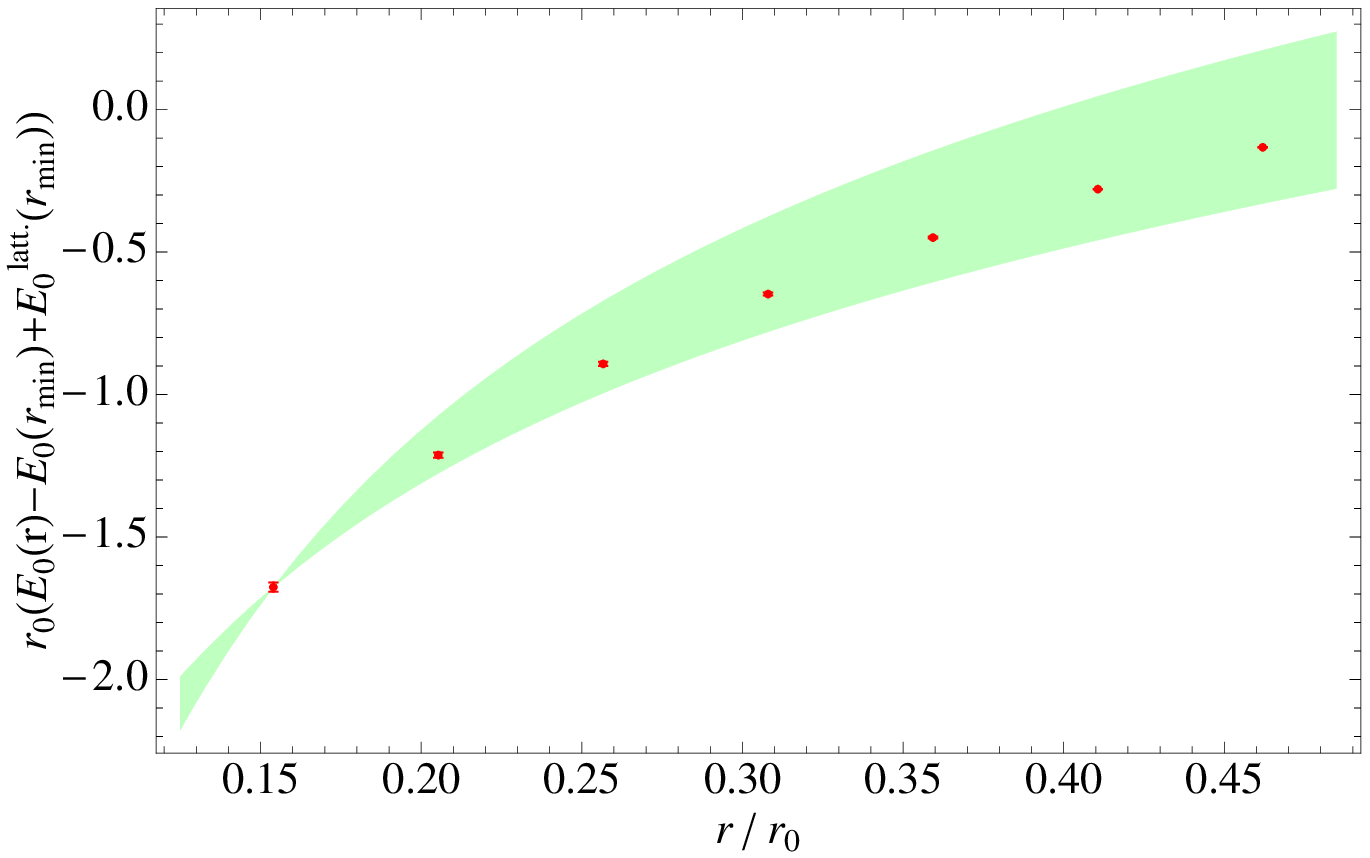}
\includegraphics[width=8cm]{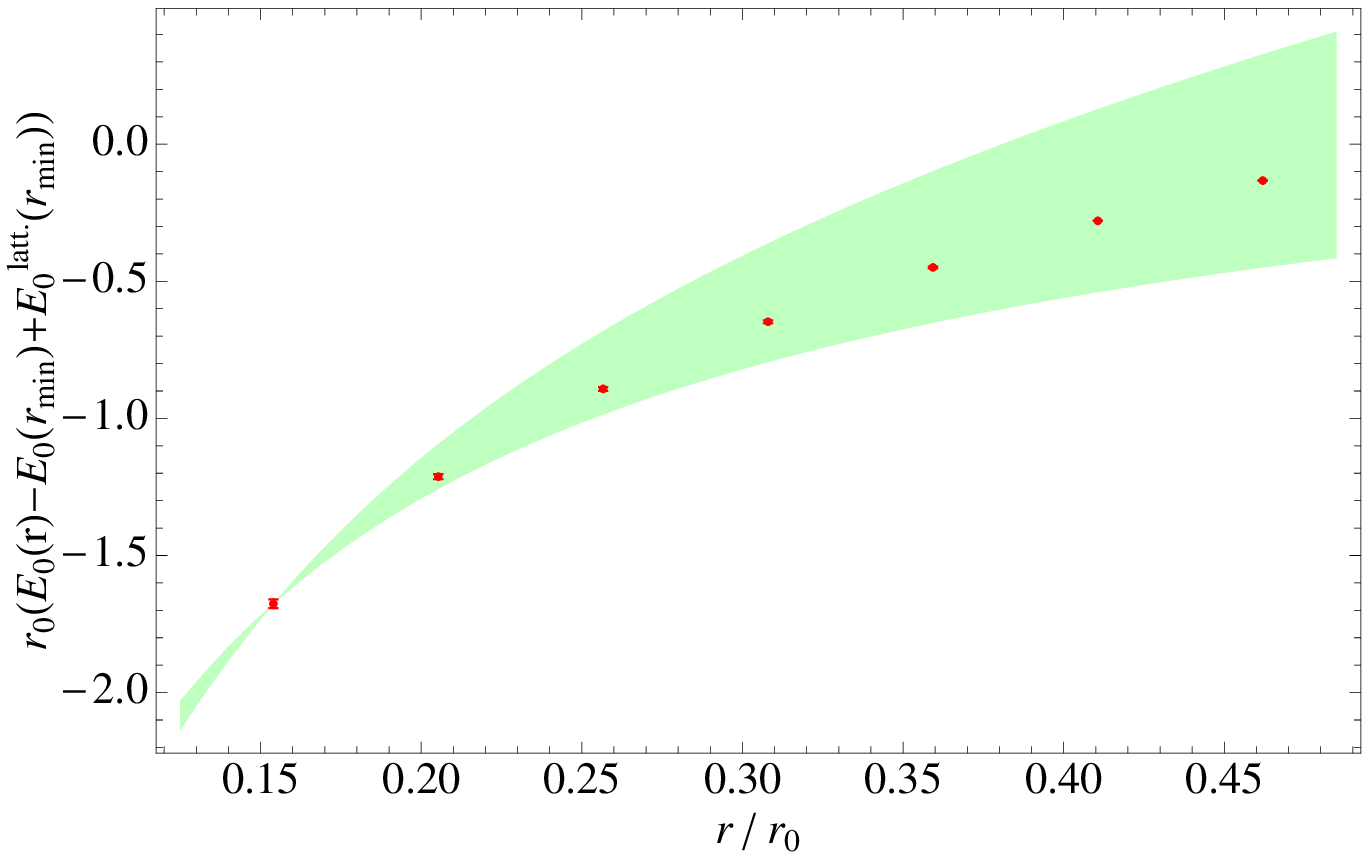}
\includegraphics[width=8cm]{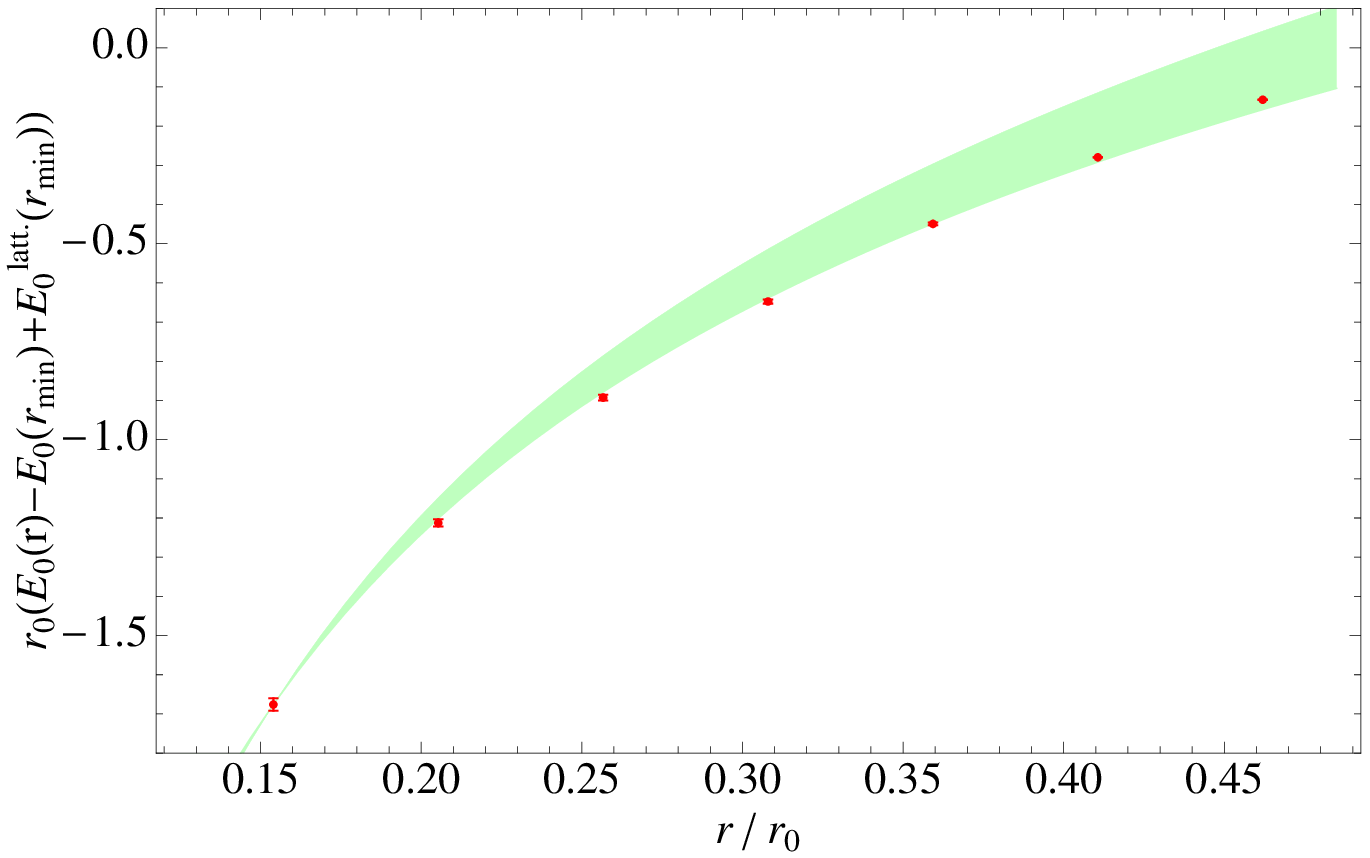}
\includegraphics[width=8cm]{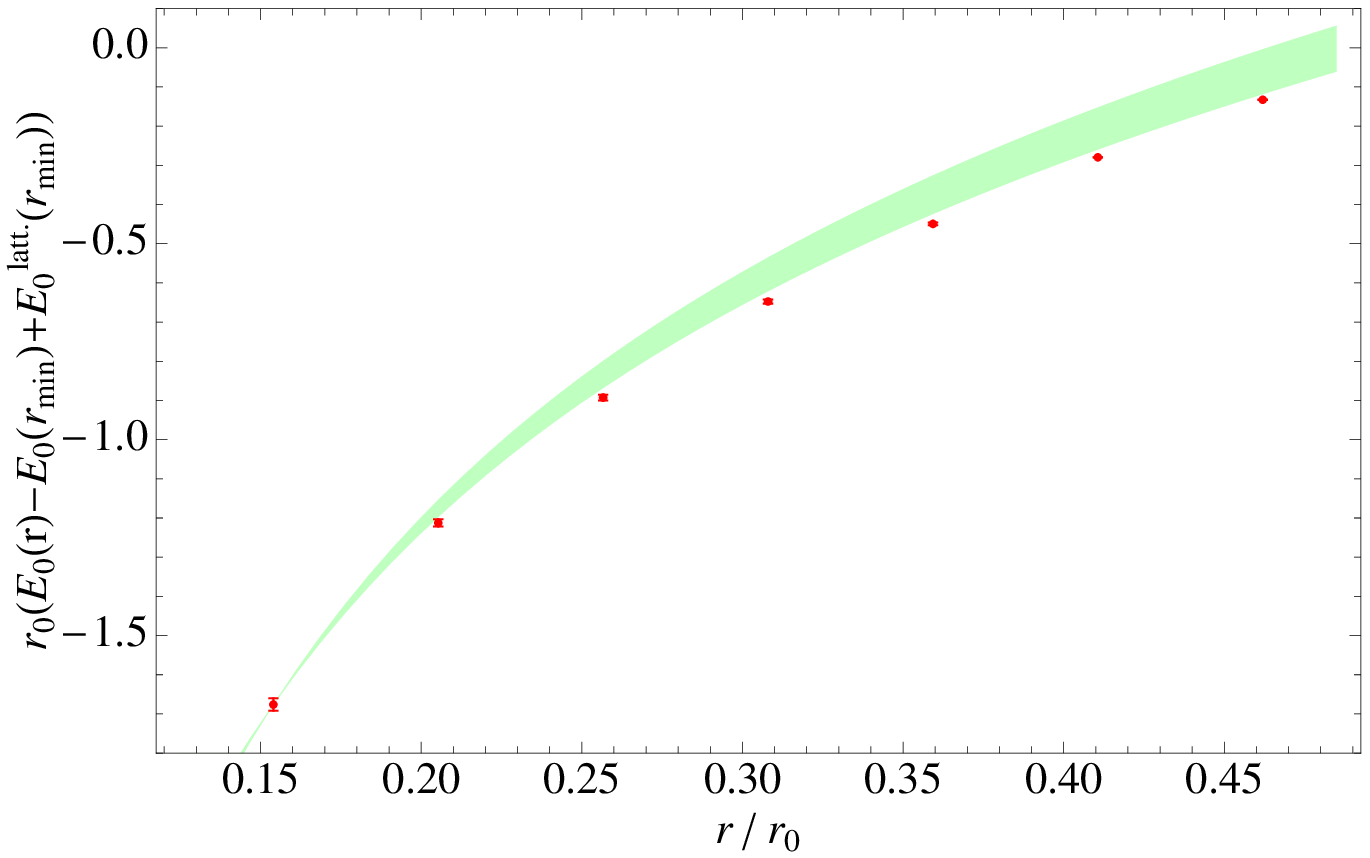}
\caption{Impact of the variation of $c_0$, $K_2$,
$\Lambda_{\MS}$ and effect of higher-order terms in the
N$^3$LL expression for the static energy respectively, 
represented as the green bands. $c_0$ has been varied by $35\%$ around
the Pad\'e value $c_0=313$. $K_2$ has been varied from -5 to
5. $\Lambda_{\MS}$ has been varied according to
$\Lambda_{\MS}\,r_0=0.602(48)$ \cite{Capitani:1998mq} and the
effect of higher-order terms has been estimated by adding the term $
\pm 10\, C_F{\als^5}/{r}$ to the N$^3$LL curve. $c_0=313$ and
$K_2=0$ have been used as central values. The red
points are the lattice data of \cite{Necco:2001xg}.}
\label{fig:varyV}
\end{figure}

According to the discussion in section \ref{subsec:scalesandpar}, we
will let $c_0$ vary by $35\%$ around the Pad\'e value $c_0=313$. $K_2$
will be varied from  -5 to 5. $\Lambda_{\MS}$ will be
varied according to the range quoted in Ref. \cite{Capitani:1998mq},
$\Lambda_{\MS}\,r_0=0.602(48)$. Finally we assess the impact of
neglecting the four-loop, $\als^5/r$, terms by simply adding the term
$\pm 10 \, C_F\als^5/r$ to the three-loop curve ($\pm 10$ is
intended to be a rough estimate of the size of the four-loop
coefficient in the RS scheme). All those variations are shown in
figure \ref{fig:varyV} as the green bands. We have used the Pad\'e
value $c_0=313$ for the plots where we vary $K_2$, $\Lambda_{\MS}$ and
the higher-order terms. $K_2=0$ is taken for the plots where we vary
$c_0$, $\Lambda_{\MS}$ and the higher-order terms. We can clearly see
that the variations due to $c_0$ and $K_2$ produce larger bands than
those due to the uncertainties in $\Lambda_{\MS}$ and higher-order
terms. Therefore, it will make sense to fit $c_0$
and $K_2$ to the lattice data. 

\subsection{Lattice comparison and extraction of $c_0$}
\label{subsec:exc0} 
\begin{figure}
\centering
\includegraphics[width=10cm]{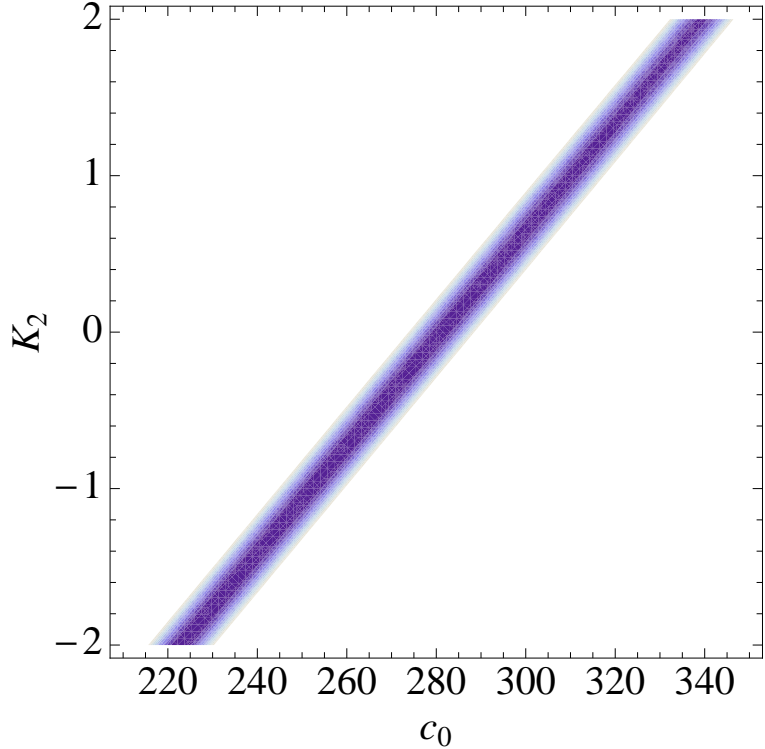}
\caption{$(\chi^2/d.o.f.)_{N^3LL}/(\chi^2/d.o.f.)_{N^2LL}$ for different values
of the $K_2$ and $c_0$ parameters. The color of each point in the
$c_0-K_2$ plane represents the value of the ratio, lighter
colors correspond to higher values and darker colors to lower
values. We show values of the ratio up to 1.}
\label{fig:scplot}
\end{figure}

\begin{figure}
\centering
\includegraphics[width=8cm]{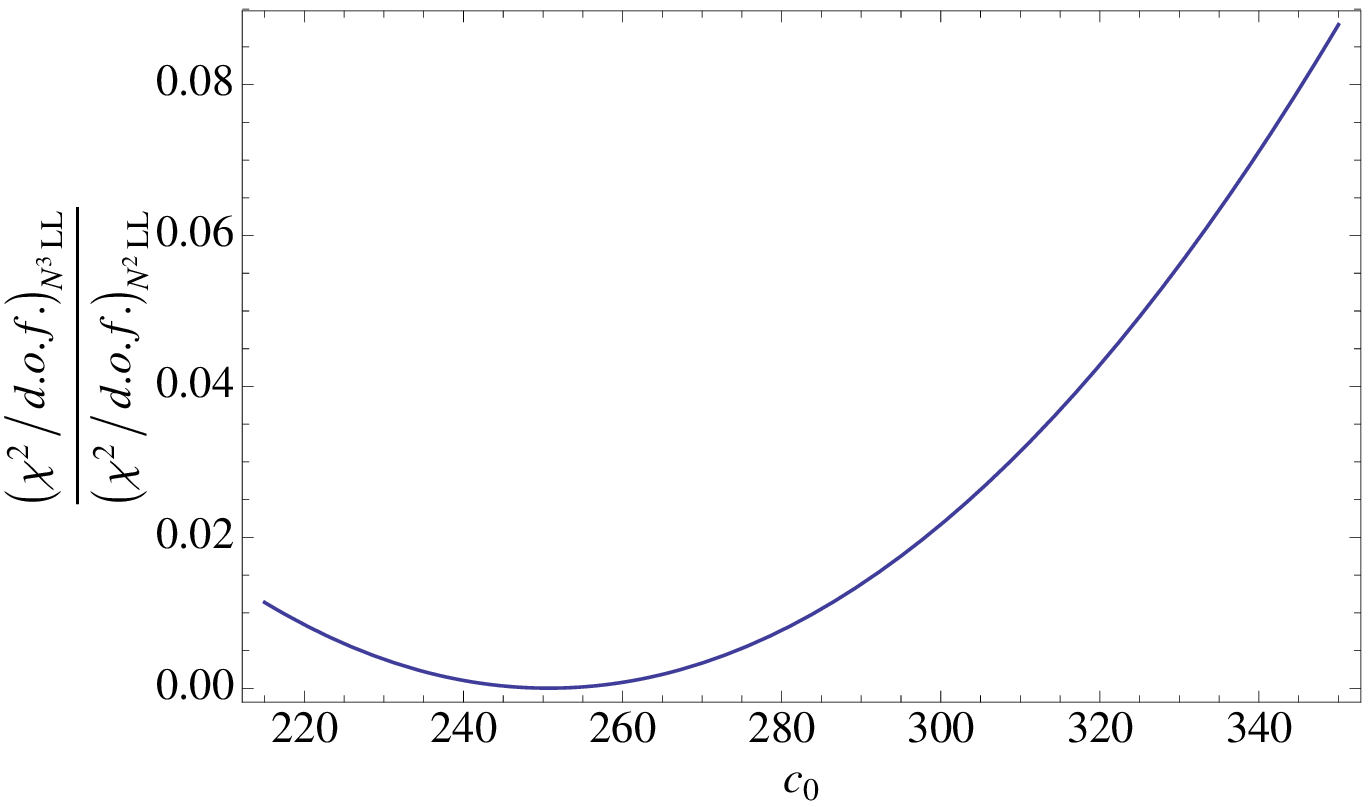}
\includegraphics[width=8cm]{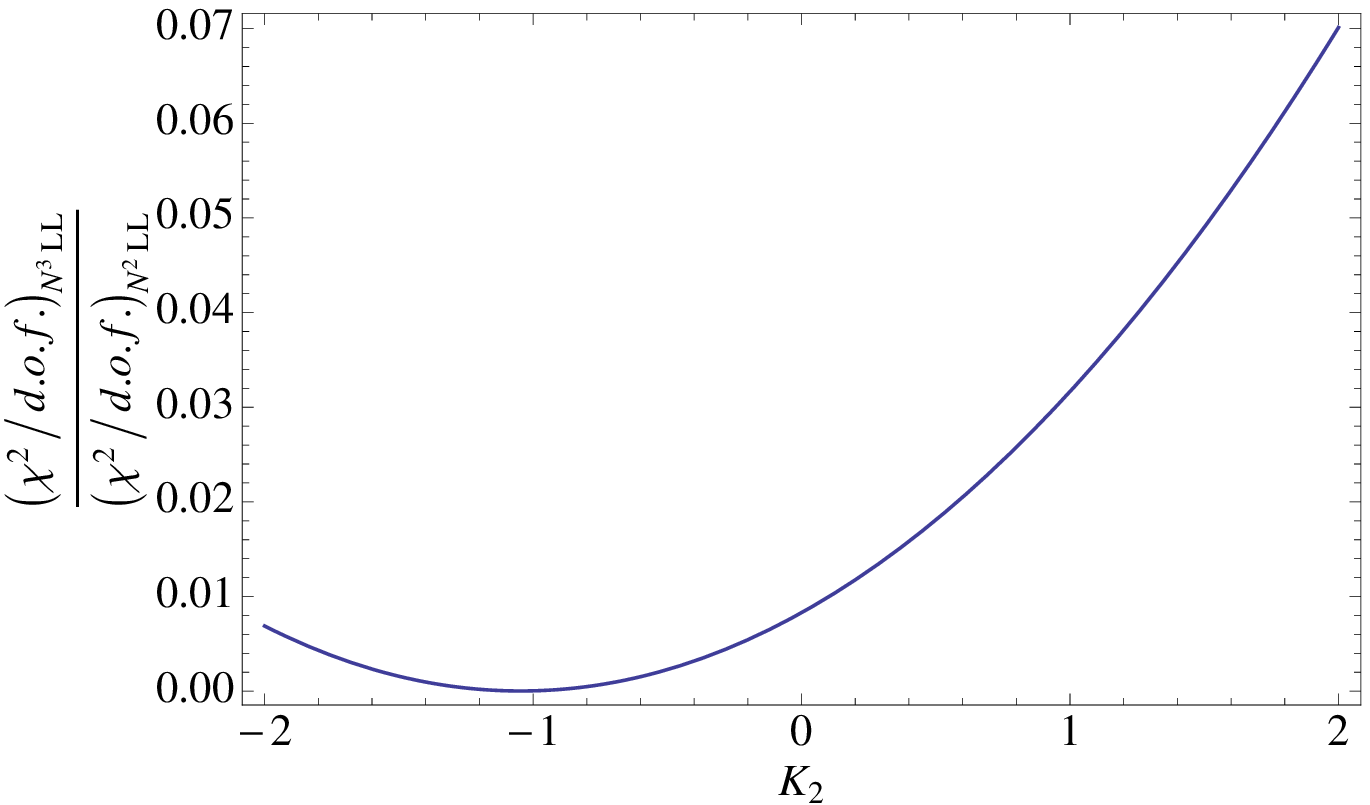}
\caption{$(\chi^2/d.o.f.)_{N^3LL}/(\chi^2/d.o.f.)_{N^2LL}$ for different values of $c_0$ (left) or $K_2$ (right), 
the other parameter is fitted.}
\label{fig:chi2plots}
\end{figure}

We are now ready to compare with lattice
data. As we have seen, the N$^3$LL expression for the static energy
depends on two unknown parameters, $K_2$ and $c_0$, which we will
obtain from a fit to the lattice data points below $r/r_0=0.5$. The
result of this two parameters fit ($\Lambda_{\MS}\,r_0=0.602$ 
and higher-order terms are set to zero) gives the values
\be
\label{eq:bestfit} 
K_2=-1.0465\,, \qquad c_0=250.7, 
\ee  
the $\chi^2/d.o.f.$ of the fit is 0.07 (we have four degrees of
freedom). Those are the best fit values, but to obtain an allowed
range of values for $c_0$, according to the lattice data, we will
analyze the $\chi^2$ of the N$^3$LL curve for different values of the
$c_0$ and $K_2$ parameters (with $\Lambda_{\MS}\,r_0=0.602$ and
higher-order terms set to zero). 
We choose the
allowed range of values for $c_0$ by identifying 
the region of the $c_0$-$K_2$ parameter space where the
reduced $\chi^2$ of the N$^3$LL curve is better than that of the
N$^2$LL curve, and, at the same time, $K_2$ retains a reasonable power
counting value. The reduced $\chi^2$ of the N$^2$LL curve is 3383 (in
this case we have six degrees of freedom). We will allow values of
$|K_2|$ up to $|K_2|=2$, to be conservatively consistent with our
power counting. Figure \ref{fig:scplot} presents a scatter plot which
shows the values of the ratio
$(\chi^2/d.o.f.)_{N^3LL}/(\chi^2/d.o.f.)_{N^2LL}$ with different
values of the $K_2$ and $c_0$ parameters. In that plot, lighter
(darker) colors correspond to higher (lower) values of the ratio. From
that we obtain the range (215,350) for $c_0$, as can be read from the
figure. To make the values of $\chi^2$ easier to visualize, we also
present, in figure \ref{fig:chi2plots}, separate scans over values of
$c_0$ and $K_2$, i.e. we scan over $c_0$ or $K_2$, fit the other
parameter for each point, and present the ratio of $\chi^2$s as a
function of $c_0$ or $K_2$. 

\begin{figure}
\centering
\includegraphics[width=12cm]{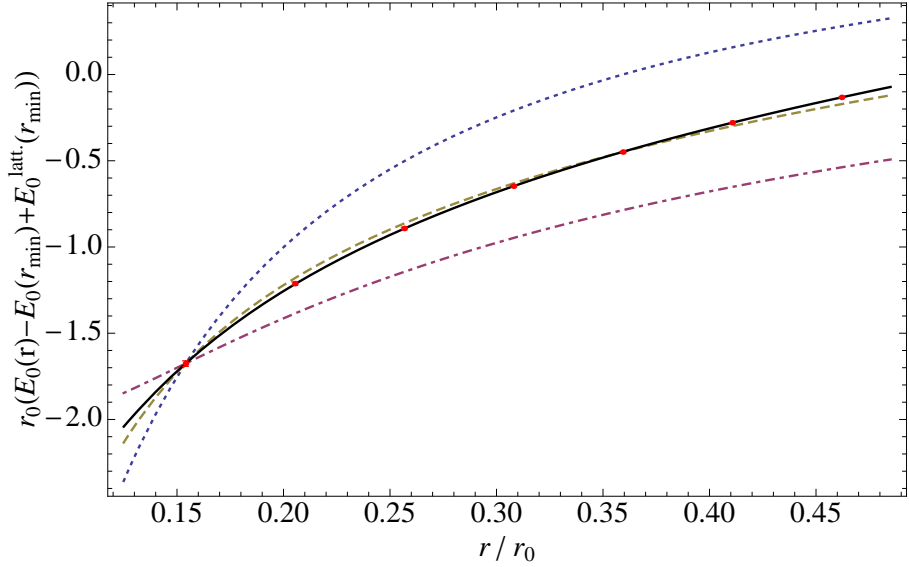}
\caption{Comparison of the singlet static energy with lattice
  data. We plot $r_o\left(E_0(r)-E_0(r_{\rm min})+E_0^{\rm latt.}(r_{\rm min})\right)$
 as a function of $r/r_0$ and the
  lattice data of \cite{Necco:2001xg} (red points). The dotted blue
  curve is at tree level, the dot-dashed magenta curve is at one loop, the
  dashed brown curve is at two-loop plus leading ultrasoft logarithmic
  resummation and the solid black curve is at three-loop plus next-to-leading 
  ultrasoft logarithmic resummation, using the best fit values of Eq. (\ref{eq:bestfit}).}
\label{fig:Vsus2ctlatt}
\end{figure}

In figure \ref{fig:Vsus2ctlatt}, we show the N$^3$LL curve,
with the best fit values for $K_2$ and $c_0$ given in (\ref{eq:bestfit}), together
with the lattice data. We also show in the plot the tree-level, one-loop 
and N$^2$LL curves for the static energy. Several comments are in
order. First, we can see that the agreement with lattice is improved
when we go from tree level, to one loop, to N$^2$LL. We also note that
the N$^3$LL curve describes very well the data. 
The fit of the N$^3$LL curve to the lattice data was not 
constrained to give a value of $K_2$ compatible with the counting, 
but this turns out to be the case (recall that our power counting requires
$|K_2|\sim\als^2/r\sim\Lambda\sim0.6$, in units of $r_0$). This gives
us much confidence in the consistency of our analysis. 
Note also that the best fit value for $c_0$ is
smaller than the Pad\'e estimate of \cite{Chishtie:2001mf}, but in
better agreement with (\ref{eq:c0from6}). We would like to emphasize
that the values of $c_0$ in Eqs. (\ref{eq:c0from6}) and
(\ref{eq:bestfit}) are obtained by completely independent
procedures. Finally, let us also note that, when we use the value of
$c_0$  in Eq. (\ref{eq:bestfit}), the convergence of the perturbative
series for the potential seems to be slightly improved, with respect to using the Pad\'e
value $c_0$=313 (see the solid black curve in figure
\ref{fig:Vsnous}).

\begin{figure}
\centering
\includegraphics[width=8cm]{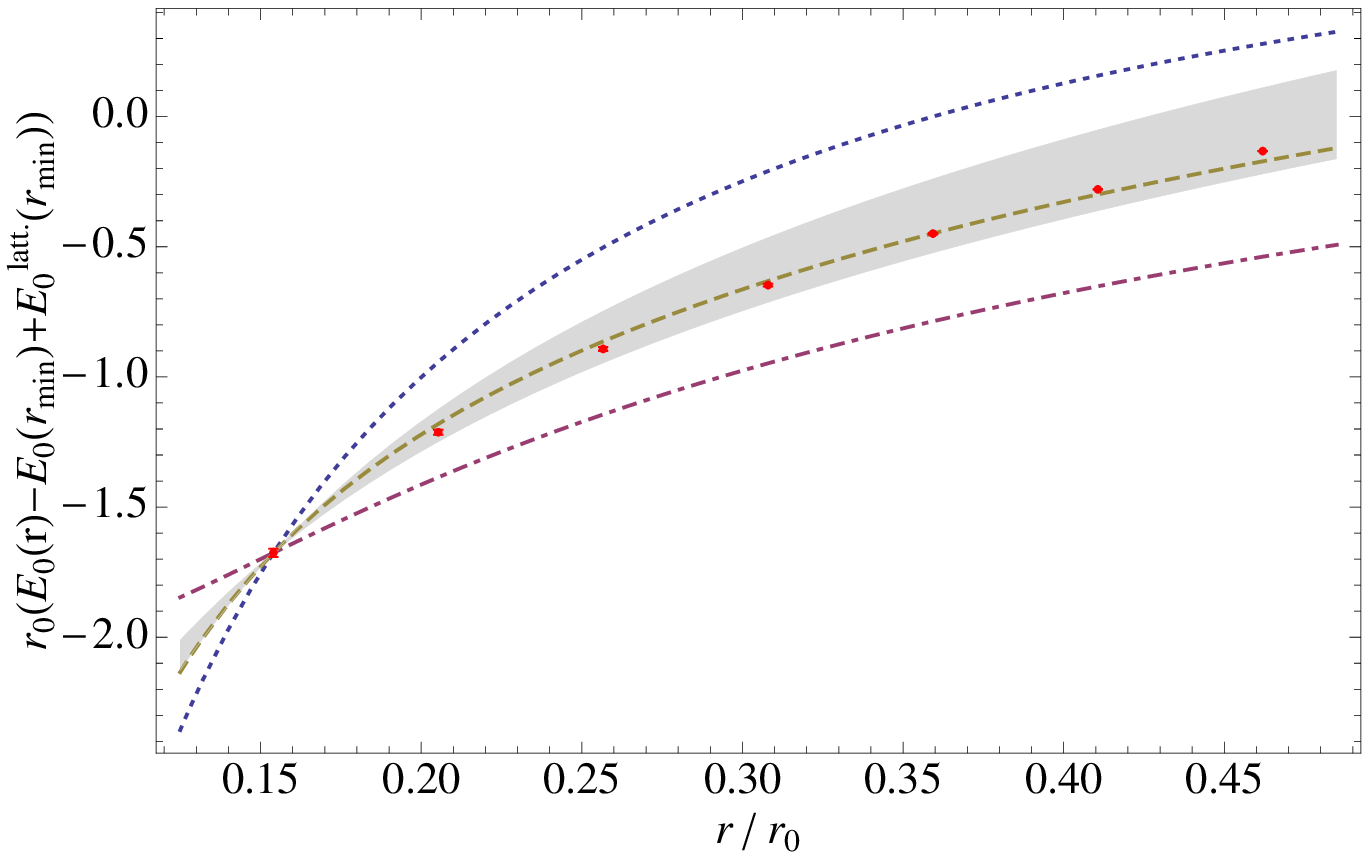}
\includegraphics[width=8cm]{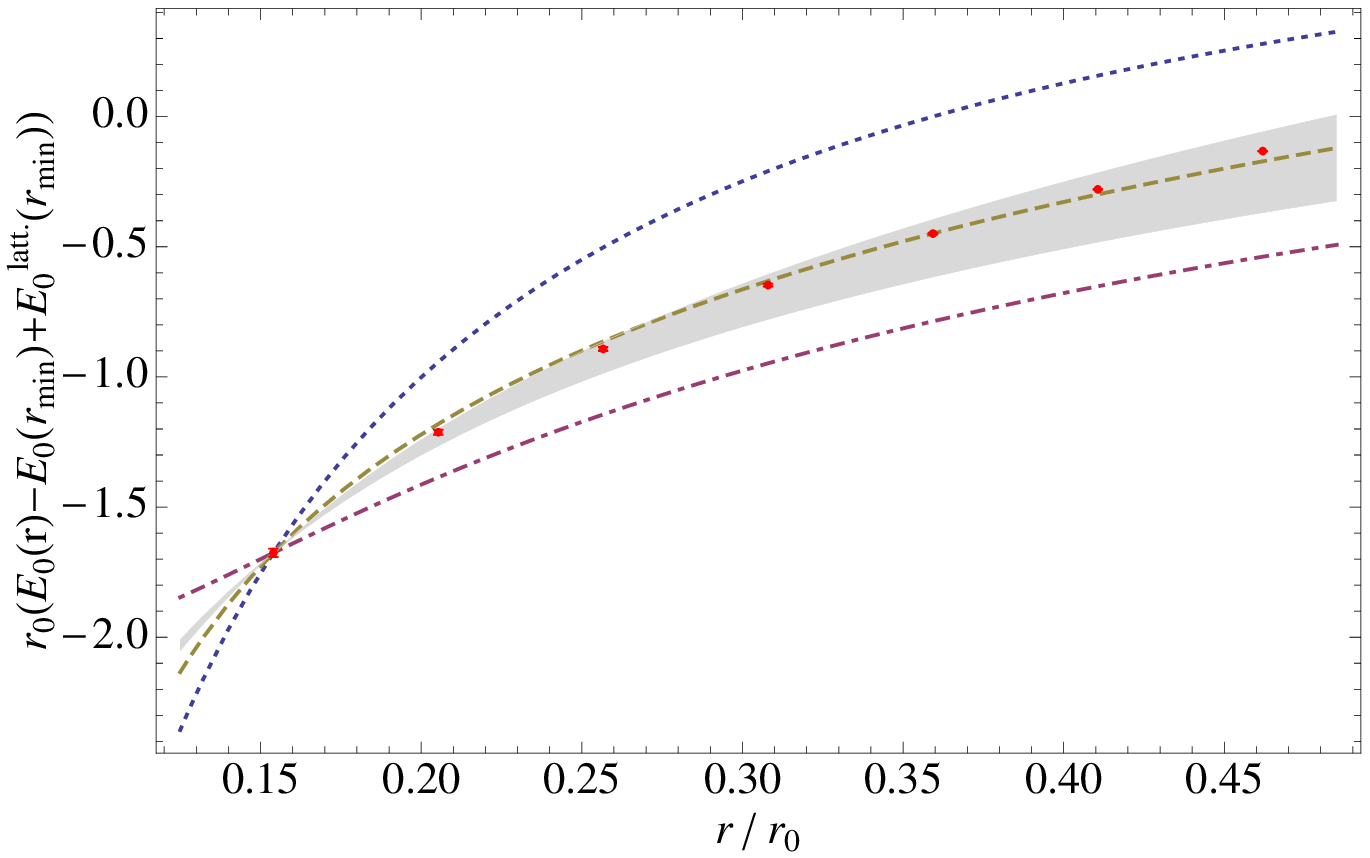}
\includegraphics[width=8cm]{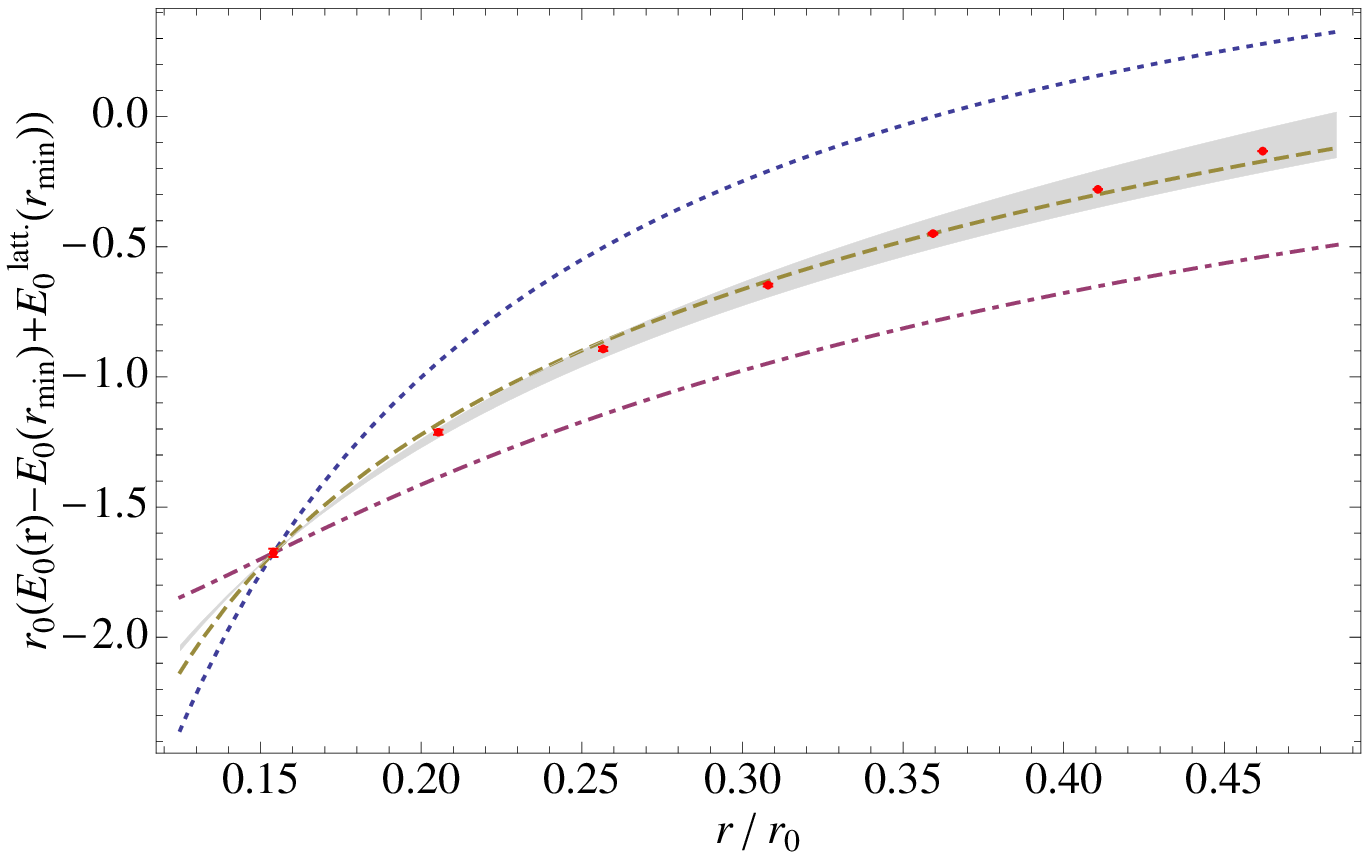}
\includegraphics[width=8cm]{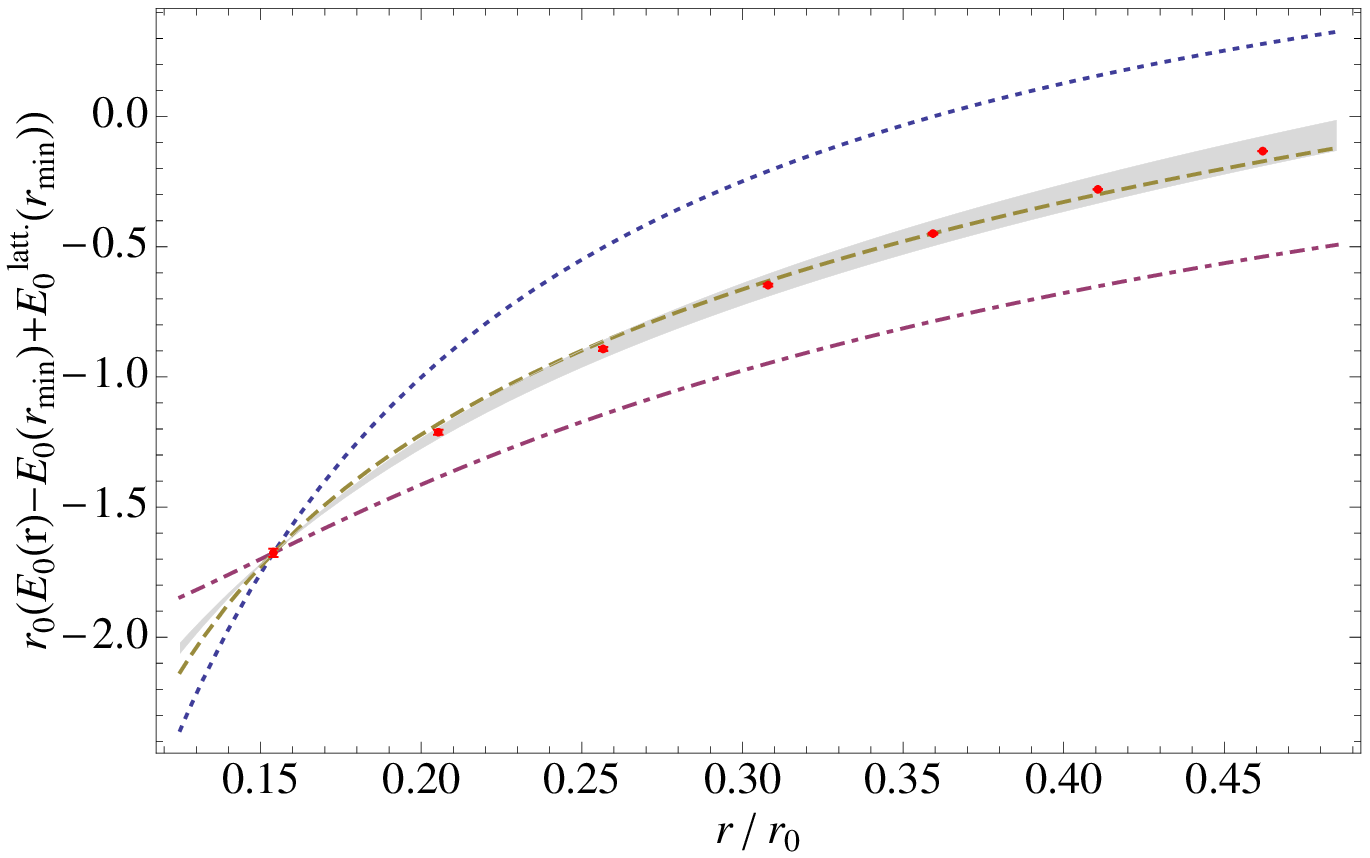}
\caption{Impact of the variation of $c_0$, $K_2$,
$\Lambda_{\MS}$ and effect of higher-order terms in the static
energy, respectively, represented as the gray bands. $c_0$ has been
varied in the interval (215, 350), $K_2$ in the interval (-2, 2),
$\Lambda_{\MS}$ has been varied according to
$\Lambda_{\MS}\,r_0=0.602(48)$ \cite{Capitani:1998mq} and the
effect of higher-order terms has been estimated by adding the term
$\pm 10 \, C_F\als^5/r$ to the N$^3$LL curve. All the bands
are obtained by fixing the other parameters at their central or best
fit values. The curves are the same as in figure
\ref{fig:Vsus2ctlatt}. The red points are the lattice data of
\cite{Necco:2001xg}.}
\label{fig:bandsE}
\end{figure}

In figure \ref{fig:bandsE} we present the bands obtained by varying
$c_0$ and $K_2$ according to the ranges described above, and the bands
induced by the variations in $\Lambda_{\MS}$ and higher-order
terms. All the bands are obtained by keeping the rest of the
parameters at their central or best fit values.

Let us comment at this point on the scheme, factorization scale and
implementation of the RS scheme dependences of the extracted value of $c_0$ (or
equivalently $\tilde{a}_{3,s}$). Concerning the scheme dependence, recall that the
scheme of Ref. \cite{Brambilla:2006wp} has been used to factorize the
ultrasoft contributions, which differs from standard 
$\MS$ scheme (we note that ultrasoft contributions were
ignored in the Pad\'e estimate of Ref. \cite{Chishtie:2001mf}). 
However, if we add to $\tilde{a}_{3,s}$ the corresponding 
non-logarithmic parts coming from $\delta_{\rm US}$,
then we obtain the non-logarithmic piece of the static energy at three
loops (which was denoted as $\tilde{a}_{3}$ in
\cite{Brambilla:2006wp}). The three-loop coefficient of the static
energy $\tilde{a}_{3}$ is independent of the scheme used to
factorize the ultrasoft contribution, as opposed to
$\tilde{a}_{3,s}$. At the practical level, the non-logarithmic part of
the ultrasoft contribution just shifts the values of $c_0$ that we quote
above by $-1\%$. Concerning the factorization scale and implementation
of the RS scheme dependences, we have checked that varying $R_s$
by $30\%$, $R_o$ by $10\%$ (according to the corresponding last known
terms of the series in (\ref{eq:RsRo})), the scales $\rho$ and $\mu$
by $10\%$ and implementing the RS subtraction with just the $d_0$ term
in (\ref{RSfullscheme}) shifts the range for $c_0$ by a maximum of $7\%$. 

We conclude by noting that both the best fit value from the
lattice comparison and Eq. (\ref{eq:c0from6}) indicate a value of
$c_0$ lower than the Pad\'e estimate $c_0=313$. The computation of
this three-loop coefficient is reported to be in progress
\cite{Smirnov:2008pn}. For the sake of comparison, the static energy
at N$^3$LO is given by
\bea 
&& \hspace{-8mm} 
E_0(r) =-\frac{C_F\als(1/r)}{r}
\Bigg\{1 + \tilde{a}_1\,\frac{\als(1/r)}{4\pi}
\nn
\\ 
&& \hspace{-6mm}  
+ \tilde{a}_{2}\,\left(\frac{\als(1/r)}{4\pi}\right)^2  
+ \left[\frac{16\,\pi^2}{3} C_A^3 \,  \ln {\frac{C_A\als(1/r)}{2}} + \tilde{a}_{3}
  \right]\! \left(\frac{\als(1/r)}{4\pi}\right)^3\Bigg\}+K_1,
\eea
with $\tilde{a}_2=\tilde{a}_{2\,s}$ and $\tilde{a_3}$ in the range $(1.08,1.17)\times 10^5$.

\section{Conclusions}
\label{sec:concl} 
We have calculated the QCD static energy at short
distances at N$^3$LL accuracy, in terms of the three-loop singlet potential,
whose coefficient $\tilde{a}_{3,s}$ for $n_f = 0$ is the only missing
ingredient in our calculation. It is remarkable that such a
higher-order calculation can be carried out analytically, which shows,
once more, what invaluable tools effective field theories provide for
higher-order calculations.  The static energy at this order turns out to depend
on two arbitrary constants, rather than one, which encode
non-perturbative effects that are competing with the weak-coupling
calculation at the considered accuracy. We have used the lattice data
of Ref.~\cite{Necco:2001xg} to extract the value of the unknown piece
of the three-loop singlet potential. Our analysis indicates the following value of
$\tilde{a}_{3,s}$:  
\be
\tilde{a}_{3,s}=1.11^{+0.06}_{-0.03}\times 10^5,
\ee
where the central value corresponds to the best fit of the N$^3$LL curve. 
For those values, an excellent agreement with lattice data is achieved in 
the region where the weak-coupling calculation is reliable. 

\bigskip

{\bf Acknowledgments}

We thank Antonio Pineda for many clarifications.
N.B., J.S. and A.V. acknowledge financial support from the MEC-INFN exchange program (Italy-Spain) 
and the RTN Flavianet MRTN-CT-2006-035482 (EU). 
J.S. acknowledge financial support from the  FPA2007-60275/ and FPA2007-66665-C02-01/
MEC grants, from the CPAN CSD2007-00042 Consolider-Ingenio
2010 program (Spain), and the 2005SGR\-00564 CIRIT grant (Catalonia). 
The work of X.G.T. was supported in part by the U.S. Department of Energy, 
Division of High Energy Physics, under contract DE-AC02-06CH11357.
The research of N.B. and A. V.  was partially supported by the DFG cluster of excellence
"Origin and structure of the universe" (www.universe-cluster.de).

\appendix

\section{Relation between $c_0$ and ${\tilde a}_{3,s}$}
\label{app:c0}
Writing the expansion of the singlet potential (for $\mu=1/r$) as 
\be
V_s=\frac{\als(1/r)}{r}\sum_{n=0}^{\infty}V_s^{(n)}\als^n(1/r),
\ee
we have
\be
\label{eq:Vs3}
V_s^{(3)}=-\frac{4}{3\pi^3}
\left(c_0+2\gamma_Ec_1+\left(4\gamma_E^2+\frac{\pi^2}{3}\right)c_2
+\left(8\gamma_E^3+2\pi^2\gamma_E+16\zeta(3)\right)c_3\right),
\ee
where for $n_f=0$
\be
\label{eq:cis}
c_1 = 290.769\,, \qquad c_2  =  \frac{1639}{16} \,, \qquad c_3  =  \frac{1331}{64}\,.
\ee
Finally, the relation between ${\tilde a}_{3,s}$ and $V_s^{(3)}$ is 
\be
- C_F \frac{{\tilde a}_{3,s}}{(4\pi)^3}=V_s^{(3)}.
\label{eq:a3Vs3}
\ee

\end{document}